\documentclass[times,twocolumn,final]{elsarticle}
\usepackage{medima}
\usepackage{framed,multirow}

\usepackage{amssymb}
\usepackage{latexsym}

\usepackage{url}
\usepackage{hyperref}
\hypersetup{pdfauthor={Name}}
\definecolor{newcolor}{rgb}{.8,.349,.1}
\journal{Medical Image Analysis}
\usepackage{tabularx}
\usepackage{threeparttable}
\usepackage{booktabs}
\usepackage{bm}
\usepackage{multirow}
\usepackage{amsmath}
\usepackage{stfloats}
\usepackage{textcomp}
\usepackage{gensymb}
\usepackage[ruled,vlined]{algorithm2e}
\usepackage{color,soul}
\usepackage{cprotect}
\usepackage{palatino}

\usepackage{scalerel}
\usepackage{tikz}
\usetikzlibrary{svg.path}
\definecolor{orcidlogocol}{HTML}{A6CE39}
\tikzset{
  orcidlogo/.pic={
    \fill[orcidlogocol] svg{M256,128c0,70.7-57.3,128-128,128C57.3,256,0,198.7,0,128C0,57.3,57.3,0,128,0C198.7,0,256,57.3,256,128z};
    \fill[white] svg{M86.3,186.2H70.9V79.1h15.4v48.4V186.2z}
                 svg{M108.9,79.1h41.6c39.6,0,57,28.3,57,53.6c0,27.5-21.5,53.6-56.8,53.6h-41.8V79.1z M124.3,172.4h24.5c34.9,0,42.9-26.5,42.9-39.7c0-21.5-13.7-39.7-43.7-39.7h-23.7V172.4z}
                 svg{M88.7,56.8c0,5.5-4.5,10.1-10.1,10.1c-5.6,0-10.1-4.6-10.1-10.1c0-5.6,4.5-10.1,10.1-10.1C84.2,46.7,88.7,51.3,88.7,56.8z};
  }
}

\newcommand\orcidicon[1]{\href{https://orcid.org/#1}{\mbox{\scalerel*{
\begin{tikzpicture}[yscale=-1,transform shape]
\pic{orcidlogo};
\end{tikzpicture}
}{|}}}}

\newcommand{\Conv}{\mathop{\scalebox{1.5}{\raisebox{-0.08ex}{$\ast$}}}}%

\begin{document}
%% -------------------------------------------------
%% TITLE, HEADER/FOOTER, ABSTRACT ------------------
%% -------------------------------------------------
\verso{A. Saha \textit{et~al.}}

\begin{frontmatter}

\title{\vspace{-6mm} \huge \textbf{End-to-end Prostate Cancer Detection in bpMRI via \\ 3D CNNs: Effects of Attention Mechanisms, Clinical Priori and Decoupled False Positive Reduction}}

\author[1]{\textbf{Anindo  \snm{Saha}}\corref{cor1}         \orcidicon{0000-0002-5091-9862}}           
\ead{anindya.shaha@radboudumc.nl}
\author[1]{\textbf{Matin   \snm{Hosseinzadeh}}\corref{cor1} \orcidicon{0000-0001-7801-2827}}
\author[1]{\textbf{Henkjan \snm{Huisman}}                   \orcidicon{0000-0001-6753-3221}} 

\cortext[cor1]{Authors with equal contribution to this research.}

\address[1]{\normalsize Diagnostic Image Analysis Group, Radboud University Medical Center, The Netherlands}

\fntext[code]{Algorithm and source code have been made publicly available at:     \\ 
\indent \enspace \href{https://grand-challenge.org/algorithms/prostate-mri-cad-cspca}{https://grand-challenge.org/algorithms/prostate-mri-cad-cspca}  \\ 
\indent \enspace \href{https://github.com/DIAGNijmegen/prostateMR\_3D-CAD-csPCa}{https://github.com/DIAGNijmegen/prostateMR\_3D-CAD-csPCa}}
\label{code}

%% \received{1 May 2013}
%% \finalform{10 May 2013}
%% accepted{13 May 2013}
%% \availableonline{15 May 2013}
%% \communicated{S. Sarkar}

\begin{abstract}
\textit{\textbf{Abstract --}}We present a multi-stage 3D computer-aided detection and diagnosis (CAD) model\hyperref[code]{$^1$} for automated localization of clinically significant prostate cancer (csPCa) in bi-parametric MR imaging (bpMRI). Deep attention mechanisms drive its detection network, targeting salient structures and highly discriminative feature dimensions across multiple resolutions. Its goal is to accurately identify csPCa lesions from indolent cancer and the wide range of benign pathology that can afflict the prostate gland. Simultaneously, a decoupled residual classifier is used to achieve consistent false positive reduction, without sacrificing high sensitivity or computational efficiency. In order to guide model generalization with domain-specific clinical knowledge, a probabilistic anatomical prior is used to encode the spatial prevalence and zonal distinction of csPCa. Using a large dataset of 1950 prostate bpMRI paired with radiologically-estimated annotations, we hypothesize that such CNN-based models can be trained to detect biopsy-confirmed malignancies in an independent cohort.

\vspace{2mm}

\noindent For 486 institutional testing scans, the 3D CAD system achieves 83.69$\pm$5.22\% and 93.19$\pm$2.96\% detection sensitivity at 0.50 and 1.46 false positive(s) per patient, respectively, with 0.882$\pm$0.030 AUROC in patient-based diagnosis --significantly outperforming four state-of-the-art baseline architectures (U-SEResNet, UNet++, nnU-Net, Attention U-Net) from recent literature. For 296 external biopsy-confirmed testing scans, the ensembled CAD system shares moderate agreement with a consensus of expert radiologists (76.69\%; \textit{kappa} $=$ 0.51$\pm$0.04) and independent pathologists (81.08\%; \textit{kappa} $=$ 0.56$\pm$0.06); demonstrating strong generalization to histologically-confirmed csPCa diagnosis.

\vspace{3mm}

\noindent \textit{\textbf{Keywords --}} prostate cancer $\hspace{0.25em} \mathbf{\cdot} \hspace{0.25em}$ magnetic resonance imaging $\hspace{0.25em} \mathbf{\cdot} \hspace{0.25em}$ convolutional neural network $\hspace{0.25em} \mathbf{\cdot} \hspace{0.25em}$ computer-aided detection and diagnosis $\hspace{0.25em} \mathbf{\cdot} \hspace{0.25em}$ anatomical prior $\hspace{0.25em} \mathbf{\cdot} \hspace{0.25em}$ deep attention

%\vspace{5mm}
%\hrule

\begin{center}

\noindent\vrule depth 0pt height 0.5pt width 0.375\textwidth\nobreak\hspace{7.5pt}\nobreak \raisebox{-3.5pt}{$\blacklozenge$}\nobreak
\hspace{7.5pt}\nobreak\vrule depth 0pt height 0.5pt width 0.375\textwidth\relax

\end{center}

\end{abstract}

\end{frontmatter}

%%\end{frontmatter}
%%% ----------------------------------------------

%% -------------------------------------------------
%% INTRODUCTION ------------------------------------
%% -------------------------------------------------
\begin{figure*}[t!]
\centering
\includegraphics[width=0.75\textwidth]{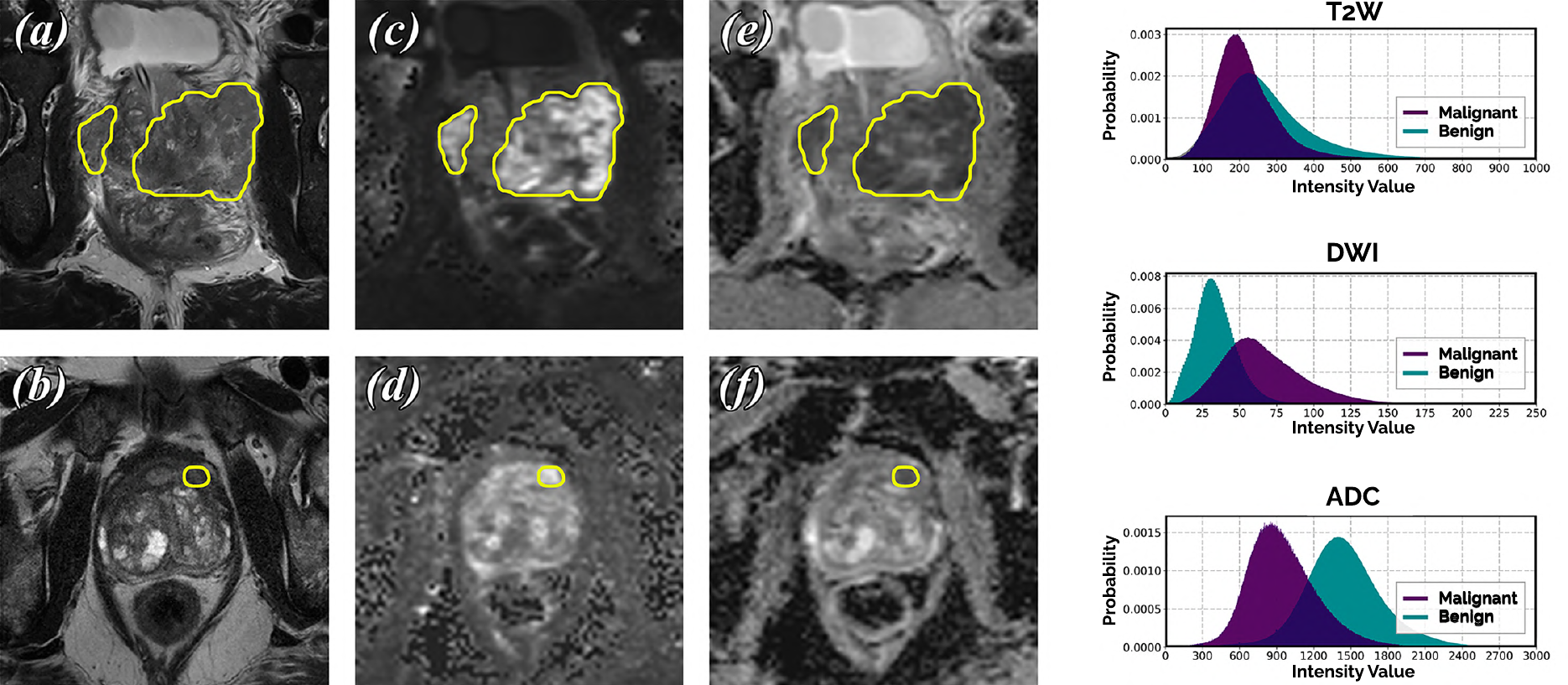}
\caption{The challenge of discriminating csPCa due to its morphological heterogeneity. \textit{\textbf{(a-b)}} T2-weighted imaging (T2W), \textit{\textbf{(c-d)}} diffusion-weighted imaging (DWI) and \textit{\textbf{(e-f)}} apparent diffusion coefficient (ADC) maps constituting the prostate bpMRI scans for two different patients are shown above, where yellow contours indicate csPCa lesions. While one of the patients had large, severe csPCa developing from both ends \textit{\textbf{(top row)}}, the other was afflicted by a single, relatively focal csPCa lesion surrounded by perceptually similar nodules of benign prostatic hyperplasia (BPH) \textit{\textbf{(bottom row)}}. Furthermore, probability density functions \textit{\textbf{(right)}} compiled from all 2732 scans used in this study, revealed a large overlap between the distributions of csPCa and non-malignant prostatic tissue across all three MRI channels.}
\label{fig1}
\end{figure*}

\section{Introduction}
\label{sec1}
Prostate cancer (PCa) is one of the most prevalent cancers in men worldwide. It is estimated that as of January, 2019, over 45\% of all men living with a history of cancer in the United States had suffered from PCa \citep{PCaStat2019}. One of the main challenges surrounding the accurate diagnosis of PCa is its broad spectrum of clinical behavior. PCa lesions can range from low-grade, benign tumors that never progress into clinically significant disease to highly aggressive, invasive malignancies, i.e. clinically significant PCa (csPCa), that can rapidly advance towards metastasis and death \citep{PCaMGMT}. In clinical practice, prostate biopsies are used to histologically assign a Gleason Score (GS) and Gleason Grade Group (GGG) to each lesion as a measure of cancer aggressiveness \citep{ISUP}. Non-targeted transrectal ultrasound (TRUS) is generally employed to guide biopsy extractions, but it remains severely prone to an underdetection of csPCa and overdiagnosis of indolent PCa \citep{BiopsySOTA}. Prostate MR imaging can compensate for these limitations of TRUS \citep{PCaMGMT, bpMRI, bpMRI_proto}, where negative MRI can rule out unnecessary biopsies by 23--45\% \citep{NewEngMRI, ZGT, UnnBiop, LancetMRI}. Prostate Imaging Reporting and Data System: Version 2 (PI-RADS v2) \citep{PIRADSv2} is the standard guideline for reading and acquiring prostate MRI, but it follows a qualitative and semi-quantitative assessment that mandates substantial expertise for proper usage. Meanwhile, csPCa can manifest as multifocal lesions of different shapes and sizes, bearing a strong resemblance to numerous non-malignant conditions (as seen in \hyperref[fig1]{Fig. 1}). In the absence of experienced radiologists, these factors can lead to low inter-reader agreement ($<$50\%) and sub-optimal interpretation \citep{LimitedPIRADS1, LimitedPIRADS3, LimitedPIRADS2, Variability}. The development of proficient and reliable csPCa detection algorithms has therefore become an important research focus in medical image computing.

\subsection{Related Work}
\label{sec1.1}
The advent of deep convolutional neural networks (CNN) has paved the way for powerful computer-aided detection and diagnosis (CAD) systems that rival human performance \citep{ISICSkin, GoogleMammo}. Machine learning models are increasingly applied for PCa detection, leveraging the high soft-tissue contrast and rich blend of anatomical and functional information present in prostate MRI. 

Multiple studies have explored architectural enhancements to extend functionality. \cite{UCLA} proposed a hybrid 2D network titled \textit{FocalNet} for joint csPCa detection and GS prediction. Over 5-fold cross-validation using 417 prostatectomy patient scans, \textit{FocalNet} achieved 0.81 AUROC in csPCa diagnosis and 87.9\% sensitivity at 1.0 false positive per patient, exclusively for MRI slices containing lesions. \cite{FPSiemens} proposed a two-stage 2D U-Net for csPCa detection, where the auxiliary second-stage module reduces false positives with contextual data from neighboring slices. Meanwhile, \cite{TwoStep} proposed a two-stage CNN model, where the first-stage segments and registers the prostate gland, while the second-stage 2D U-Net performs binary classification of csPCa (0.86 pixel-level AUROC on 20 unseen histologically-confirmed cases). Alternatively, \cite{CorrSig} proposed a two-stage CNN model, which extracts correlated cancer signatures from prostate MRI combined with its whole-mount histopathology mapping (0.86 pixel-level AUROC on 20 unseen prostatectomy patients).

Cancerous lesions stemming from the prostatic peripheral zone (PZ) exhibit different morphology and pathology than those developing from the transitional zone (TZ) \citep{csPCaZones, PIRADSv2, bpMRI}. \cite{ZonalPrior} highlights the merits of utilizing this priori through an early fusion of probabilistic zonal segmentations inside a 2D CAD system --demonstrating how the inclusion of PZ and TZ segmentations can introduce an average increase of 5.3\% detection sensitivity, between 0.5--2.0 false positives per patient. In a separate study, \cite{WeakProstate} constructed a probabilistic 2D prevalence map from 1055 MRI slices. Depicting the typical sizes, shapes and locations of malignancy across the prostate anatomy, this map was used to weakly supervise a 2D U-Net for PCa detection. Both methods underline the value of clinical priori and anatomical features --factors known to play an equally important role in radiomics-based solutions \citep{MLAnatoFeat02, MLAnatoFeat01}.

The vast majority of CAD systems for csPCa operate solely on a 2D-basis, citing computational limitations and the non-isotropic imaging protocol of prostate MRI as their primary rationale. \cite{UoT_PCa_CLF} tackled this challenge by employing dedicated 2D ResNets for each slice in a patient scan and aggregating all slice-level predictions with a Random Forest classifier (0.84 patient-level AUROC on 108 unseen biopsy-confirmed cases). \cite{Patch3D2020} proposed a patch-based approach, passing highly-localized regions of interest (ROI) through a standard 3D CNN (0.90 AUROC over 8-fold cross-validation using 200 biopsy-confirmed patients). \cite{ECCV_2.5D} followed a 2.5D approach as a compromise solution, sacrificing the ability to harness multiple MRI channels for an additional pseudo-spatial dimension.

In recent years, a number of retrospective studies have investigated the growing potential of CAD systems relative to radiologists. \cite{SanfordDLPR} compared the PI-RADS classification performance of a four-class 2D ResNet with expert radiologists, reaching 56\% agreement on 68 testing scans. \cite{RadPRvsDL} used an ensemble of 2D U-Nets to achieve statistically similar csPCa detection performance as a cohort of trained radiologists on 62 biopsy-confirmed testing scans. \cite{SPCNet2021} developed a 2.5D HED architecture, reaching 0.75 and 0.80 lesion-level AUROC on 293 biopsy-confirmed and 23 prostatectomy patient cases, respectively --approaching radiologists performance. Studies have also investigated the robustness of CAD systems across multi-institutional, multi-vendor testing sets \citep{ErasmusMC}, reaching 0.93 AUROC on 40 biopsy-confirmed and 7 prostatectomy patient cases collected across five institutions \citep{MultiInst}.

\subsection{Contributions}
\label{sec1.2}
In this research, we harmonize several state-of-the-art techniques from recent literature to present a novel end-to-end 3D CAD system that generates voxel-level detections of csPCa in prostate MRI. Key contributions of our study are, as follows:

\begin{itemize}
    \item We examine a detection network with dual-attention mechanisms, which can adaptively target highly discriminative feature dimensions and salient prostatic structures in bpMRI, across multiple resolutions, to reach peak detection sensitivity at lower false positive rates.
    \item We study the effect of employing a residual patch-wise 3D classifier for decoupled false positive reduction and we investigate its utility in improving baseline specificity, without sacrificing high detection sensitivity. 
    \item We develop a probabilistic anatomical prior, capturing the spatial prevalence and zonal distinction of csPCa from a large training dataset of 1584 MRI scans. We investigate the impact of encoding the computed prior into our CNN architecture and we evaluate its ability to guide model generalization with domain-specific clinical knowledge.
    \item We hypothesize that our model can train on radiologically-estimated annotations, begin to generalize beyond, and in turn, accurately detect histologically-confirmed malignancies, given a large training cohort with rich information. We evaluate performance across large, multi-institutional testing datasets: 486 institutional and 296 independent patient scans annotated using PI-RADS v2 and histological grades, respectively. Our benchmark includes a consensus score of expert radiologists to assess clinical viability.
\end{itemize}
%%% ----------------------------------------------

%% -------------------------------------------------
%% MATERIALS & METHODS -----------------------------
%% -------------------------------------------------
\section{Material and Methods}
\label{sec2}
\subsection{Dataset}
\label{sec2.1}
The primary dataset was a cohort of 2436 consecutive prostate MRI studies (2317 patients) from Radboud University Medical Center (RUMC), acquired over the period January, 2016--January, 2018. All cases were paired with radiologically-estimated annotations of csPCa derived via PI-RADS v2. From here, 1584 (65\%), 366 (15\%) and 486 (20\%) scans were split into training, validation and testing (TS1) sets, respectively, via double-stratified sampling --preserving the same class balance (\textit{benign} or \textit{malignant}) while ensuring non-overlapping patients, between each subset of data. Additionally, 296 prostate MRI scans (296 patients) from Ziekenhuisgroep Twente (ZGT), acquired over the period March, 2015--January, 2017, were used to curate an external testing set (TS2). TS2 annotations included biopsy-confirmed histological grades (GS, GGG).

\subsubsection{Bi-Parametric MRI Scans}
\label{sec2.1.1}
Patients were biopsy-naive men (RUMC: \{median age: 66 yrs, IQR: 61--70\}, ZGT: \{median age: 65 yrs, IQR: 59--68\}) with elevated levels of PSA (RUMC: \{median level: 8 ng/mL, IQR: 5--11\}, ZGT: \{median level: 6.6 ng/mL, IQR: 5.1--8.7\}). Imaging was performed on 3T MR scanners with surface coils (RUMC: \{89.9\% on Magnetom Trio/Skyra, 10.1\% on Prisma\}, ZGT: \{100\% on Skyra\}; Siemens Healthineers, Erlangen). Acquisitions were obtained following standard mpMRI protocols in compliance with PI-RADS v2 \citep{bpMRI_proto}. Given the limited role of dynamic contrast-enhanced (DCE) imaging in mpMRI, in recent years, bpMRI has emerged as a practical alternative --achieving similar performance, while saving time and the use of contrast agents \citep{PIRADS_2019, DCEUtil, Systematic_2020}. Similarly, in this study, we used bpMRI sequences only, which included T2-weighted (T2W) and diffusion-weighted imaging (DWI). Apparent diffusion coefficient (ADC) maps and high b-value DWI (b$\geq$1400 s/mm$^2$) were computed from the raw DWI scans. Due to the standardized precautionary measures (e.g. minimal temporal difference between acquisitions, administration of antispasmodic agents to reduce bowel motility, use of rectal catheter to minimize distension, etc.) \citep{bpMRI_proto} taken in the imaging protocol, we observed negligible patient motion across the different sequences. Thus, no additional registration techniques were applied, in agreement with clinical recommendations \citep{ISUP} and recent studies \citep{UCLA}. Further details on patient demographics, study inclusion/exclusion criteria and acquisition parameters can be found in the \textit{Supplementary Materials}.

\subsubsection{Clinical Annotations}
\label{sec2.1.2}
All patient cases were read during regular clinical routine via PI-RADS v2. At RUMC, all cases were read by at least one of six radiologists (4--25 years of experience) and difficult cases were jointly examined with an expert radiologist (25 years of experience with prostate MRI). At ZGT, all cases were read by two radiologists (6, 24 years of experience) and independently reviewed by two expert radiologists (5, 25 years of experience with prostate MRI) in consensus. In this study, we flagged any detected lesions marked PI-RADS 4 or 5 as csPCa$^{(\textit{PR})}$. All patients at ZGT underwent TRUS-guided biopsies performed by a urologist, blinded to the imaging results. In the presence of any suspicious lesions (PI-RADS$>2$), patients also underwent in-bore MRI-guided biopsies. All tissue samples were graded by general pathologists \citep[as detailed in][]{ZGT} and independently reviewed by an experienced uropathologist (25 years of experience), where cores containing cancer were assigned GS and GGG \citep[in compliance with][]{ISUP}. Any lesion graded GS $>$ 3+3 or GGG $>$ 1 was marked as csPCa$^{(\textit{GS})}$. All instances of csPCa$^{(\textit{PR})}$ and csPCa$^{(\textit{GS})}$ were carefully delineated on a voxel-level basis by trained students (6--18 months of expertise), under the supervision of an experienced radiologist (7 years of experience). 

Upon complete annotation, the RUMC and ZGT datasets contained 1527 and 210 \textit{benign} cases, along with 909 and 86 \textit{malignant} cases ($\geq 1$ csPCa lesion), respectively. Moreover, on a lesion-level basis, the RUMC dataset contained 1092 csPCa$^{(\textit{PR})}$ lesions (mean frequency: 1.21 lesions per \textit{malignant} scan; median size: 1.05 cm$^3$, range: 0.01--61.49 cm$^3$), while the ZGT dataset contained 97 csPCa$^{(\textit{GS})}$ lesions (mean frequency: 1.05 lesions per \textit{malignant} scan; median size: 1.69 cm$^3$, range: 0.23--22.61 cm$^3$). Further details on the distribution of PI-RADS and GGG scores across the study population can be found in the \textit{Supplementary Materials}.

\begin{figure*}[t!]
\centering
\includegraphics[width=\textwidth]{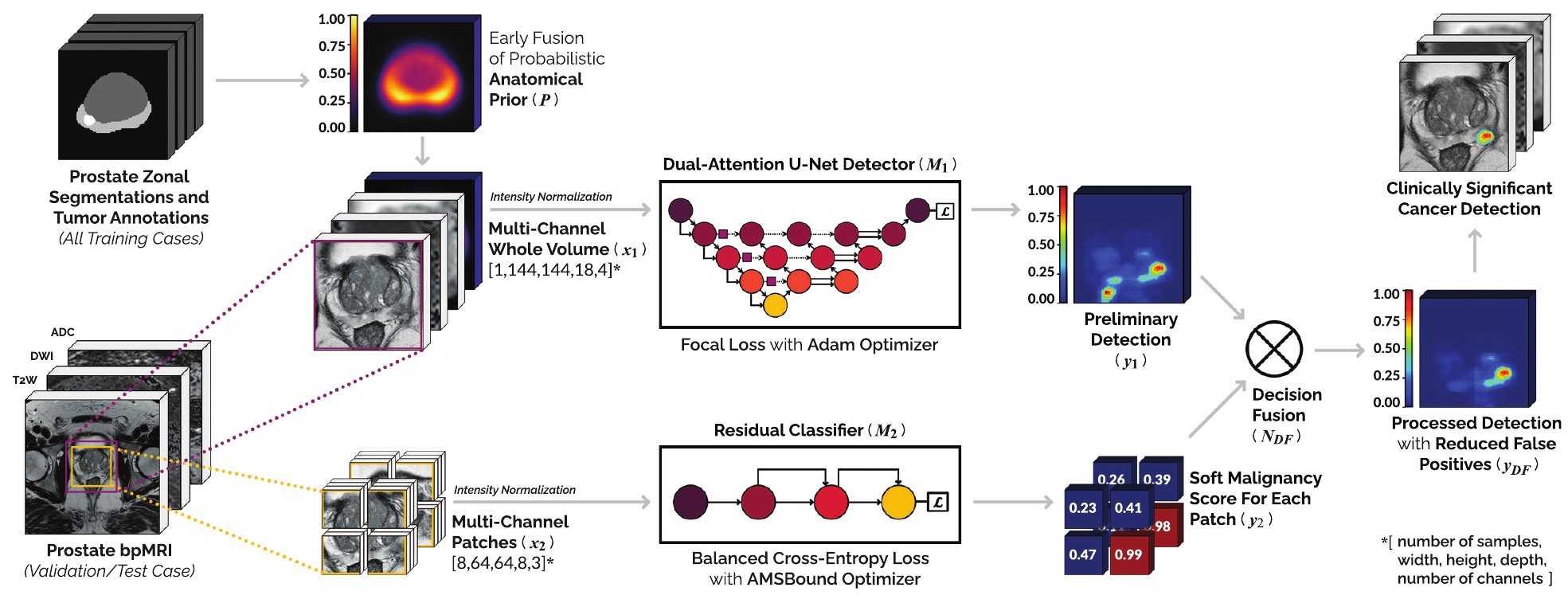}
\caption{Proposed end-to-end framework for computing voxel-level detections of csPCa in validation/test samples of prostate bpMRI. The model center-crops two ROIs from the multi-channel concatenation of the patient's T2W, DWI and ADC scans for the input of its detection and classification 3D CNN sub-models ($M_1$, $M_2$). $M_1$ leverages an anatomical prior $P$ in its input $x_1$ to synthesize spatial priori and generate a preliminary detection $y_1$. $M_2$ infers on a set of overlapping patches $x_2$ and maps them to a set of probabilistic malignancy scores $y_2$. Decision fusion node $N_{DF}$ aggregates $y_1$, $y_2$ to produce the model output $y_{DF}$ in the form of a post-processed csPCa detection map with high sensitivity and reduced false positives.}
\label{fig2}
\end{figure*}

\subsubsection{Prostate Zonal Segmentations}
\label{sec2.1.3}
Multi-class segmentations of prostatic TZ and PZ were generated for each scan using a multi-planar, anisotropic 3D U-Net from a separate study \citep{3DSegMod}. Trained using a subset of 40 patient scans from the RUMC training cohort, the network achieved an average Dice Similarity Coefficient (DSC) of 0.90$\pm$0.01, 0.85$\pm$0.02 and 0.63$\pm$0.03 for whole-gland, TZ and PZ segmentation, respectively, over 5$\times$5 nested cross-validation \citep{NestedCV}. We used these zonal segmentations to construct and apply the anatomical prior (as detailed in \hyperref[sec2.2.3]{Section 2.2.3}). For this study, the goal of the zonal segmentations was to establish object-level, prior-to-image correspondence, rather than voxel-level matching. Thus, high quality segmentations with precise contour definitions were not mandatory.

\subsection{Model Architecture}
\label{sec2.2}
The architecture of our proposed CAD solution comprises of two parallel 3D CNNs ($M_1$, $M_2$) followed by a decision fusion node $N_{DF}$, as shown in \hyperref[fig2]{Fig. 2}. Based on our observations in previous work \citep{ZonalPrior, 3DSegMod}, we opted for anisotropically-strided 3D convolutions in both $M_1$ and $M_2$ to process the bpMRI data, which resemble multi-channel stacks of 2D images rather than full 3D volumes.  Prior to usage, all acquisitions were spatially resampled to a common axial in-plane resolution of 0.5 mm$^2$ and slice thickness of 3.6 mm via B-spline interpolation. T2W and DWI channels were normalized to zero mean and unit standard deviation, while ADC channels were linearly normalized from [0,3000] to [0,1] in order to retain their clinically relevant numerical significance \citep{bpMRI}. Anatomical prior $P$, constructed using the prostate zonal segmentations and csPCa$^{(\textit{PR})}$ annotations in the training dataset, is encoded in $M_1$ to infuse spatial priori. At train-time, $M_1$ and $M_2$ are independently optimized using different loss functions and target labels. At test-time, $N_{DF}$ is used to aggregate their predictions ($y_1$, $y_2$) into a single output detection map $y_{DF}$.

\subsubsection{Detection Network}
\label{sec2.2.1}
\begin{figure*}[b!]
\centering
\includegraphics[width=1.00\textwidth]{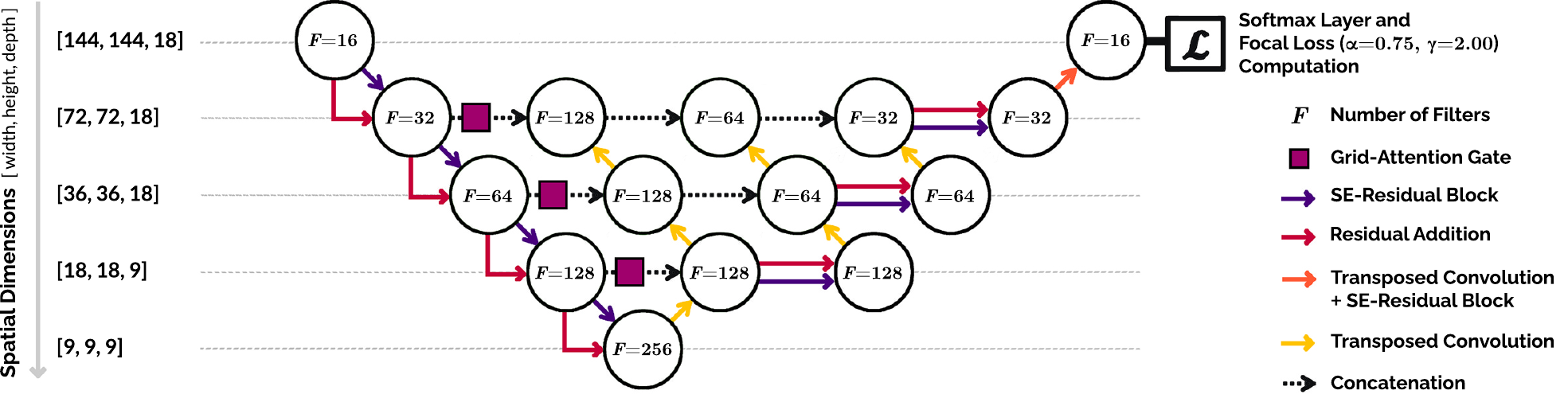}
\caption{Architecture schematic for the Dual-Attention U-Net ($M_1$). $M_1$ is a modified adaptation of the UNet++ architecture \citep{UNet++}, utilizing a pre-activation residual backbone \citep{ResNetv2} with \textit{Squeeze-and-Excitation} (SE) channel-wise attention mechanism \citep{SE} and grid-attention gates \citep{GridAtn-Gates}. All convolutional layers in the encoder and decoder stages are activated by ReLU and LeakyReLU, respectively, and use kernels of size $3\times3\times3$ with $L_2$ regularization ($\beta=0.001$). Both downsampling and upsampling operations throughout the network are performed via anisotropic strides. Dropout nodes ($rate=0.50$) are connected at each scale of the decoder to alleviate train-time overfitting.}
\label{fig3}
\end{figure*}

The principal component of our proposed model is the dual-attention detection network or $M_1$, as shown in \hyperref[fig2]{Fig. 2}, \hyperref[fig3]{3}. It is used to generate the preliminary voxel-level detection of csPCa in prostate bpMRI scans with high sensitivity. Typically, a prostate gland occupies 45--50 cm$^3$, but it can be significantly enlarged in older males and patients afflicted by BPH \citep{ProstateVolume}. The input ROI of $M_1$, measuring 144$\times$144$\times$18 voxels per channel or nearly 336 cm$^3$, includes and extends well beyond this window to utilize surrounding peripheral and global anatomical information. $M_1$ trains on whole-image volumes equivalent to its total ROI, paired with fully delineated annotations of csPCa$^{(\textit{PR})}$ as target labels. Since the larger ROI and voxel-level labels contribute to a severe class imbalance (1:153) at train-time, we use a focal loss function to train $M_1$. Focal loss addresses extreme class imbalance in one-stage dense detectors by weighting the contribution of easy to hard examples, alongside conventional class-weighting \citep{FocalLoss}. In a similar study for joint csPCa detection in prostate MRI, the authors credited focal loss as one of the pivotal enhancements that enabled their CNN solution, titled \textit{FocalNet} \citep{UCLA}. 

For an input volume, $x_1$ $=$ $({x_1}^1$, ${x_1}^2$,$..., {x_1}^n)$ derived from a given scan, let us define its target label $Y_1$ $=$ $({Y_1}^1$, ${Y_1}^2$,$..., {Y_1}^n)$ $\in$ $\{0,1\}$, where $n$ represents the total number of voxels in $x_1$. We can formulate the focal loss function of $M_1$ for a single voxel in each scan, as follows:

{\setlength{\abovedisplayskip}{-3pt}
\setlength{\belowdisplayskip}{0pt}
\begin{equation*}
\begin{split}
FL({x_1}^i,{Y_1}^i) = &-\alpha(1-{y_1}^i)^\gamma {Y_1}^i\text{log}{y_1}^i  \\ 
                      &-(1-\alpha)({y_1}^i)^\gamma (1-{Y_1}^i)\text{log}(1-{y_1}^i) \indent i \in [1,n]
\end{split}
\end{equation*}}

\noindent Here, ${y_1}^i = p(O\text{=}1|{x_1}^i) \in [0,1]$, represents the probability of ${x_1}^i$ being a \textit{malignant} tissue voxel as predicted by $M_1$, while $\alpha$ and $\gamma$ represent weighting hyperparameters of the focal loss. At test-time, $y_1 = ({y_1}^1, {y_1}^2,..., {y_1}^n) \in [0,1]$, i.e. a voxel-level, probabilistic csPCa detection map for $x_1$, serves as the final output of $M_1$ for each scan.

We choose 3D U-Net \citep{U-Net, 3DUNet} as the base architecture of $M_1$, for its ability to summarize multi-resolution, global anatomical features \citep{AnatomicalPriors, nnUNet} and generate an output detection map with voxel-level precision. Pre-activation residual blocks \citep{ResNetv2} are used at each scale of $M_1$ for deep feature extraction. Architecture of the decoder stage is adapted into that of a modified UNet++ \citep{UNet++} for improved feature aggregation. UNet++ uses redesigned encoder-decoder skip connections that implicitly enable a nested ensemble configuration. In our adaptation, its characteristic property of feature fusion from multiple semantic scales is used to achieve similar performance, while dense blocks and deep supervision from the original design are forgone to remain computationally lightweight. 

Two types of differentiable, soft attention mechanisms are employed in $M_1$ to highlight salient information throughout the training process, without any additional supervision. Channel-wise \textit{Squeeze-and-Excitation} (SE) attention \citep{SE, USENet} is used to amplify the most discriminative feature dimensions at each resolution. Grid-attention gates \citep{GridAtn-Gates} are used to automatically learn spatially important prostatic structures of varying shapes and sizes. While the former is integrated into every residual block to guide feature extraction, the latter is placed at the start of skip-connections to filter the semantic features being passed onto the decoder. During backpropagation, both attention mechanisms work collectively to suppress gradients originating from background voxels and inessential feature maps. Similar combinations of dual-attention mechanisms have reached state-of-the-art performance in semantic segmentation challenges \citep{DualAttention} and PCa diagnosis \citep{Panoptic}, sharing an ability to integrate local features with their global dependencies.

\vspace{2mm}
\subsubsection{Classifier for False Positive Reduction}
\label{sec2.2.2}
The goal of the classification network, $M_2$, is to improve overall model specificity via independent, binary classification of each scan and its constituent segments. It is effectuated by $N_{DF}$, which factors in these predictions from $M_2$ to locate and penalize potential false positives in the output of $M_1$. $M_2$ has an input ROI of 112$\times$112$\times$12 voxels per channel or nearly 136 cm$^3$, tightly centered around the prostate. While training on the full ROI volume has the advantage of exploiting extensive spatial context, it results in limited supervision by the usage of a single coarse, binary label per scan. Thus, we propose patch-wise training using multiple, localized labels, to enforce fully supervised learning. We define an effective patch extraction policy as one that samples regularly across the ROI to densely cover all spatial positions. Sampled patches must also be large enough to include a sufficient amount of context for subsequent feature extraction. Random sampling within a small window, using the aforementioned criteria, poses the risk of generating highly overlapping, redundant training samples. However, a minimum level of overlap can be crucial, benefiting regions that are harder to predict by correlating semantic features from different surrounding context \citep{SpatialContext}. To facilitate these conditions, the ROI is uniformly divided into a set of eight octant training samples $x_2$, measuring $64\times64\times8$ voxels each, with an overlap stride of 16 and 4 voxels across their in-plane and through-plane dimensions, respectively. As a result, training samples consistently and deterministically cover the total ROI, while 71.85\% of each patch overlaps with a portion of its seven neighboring patches. 

For input patches, $x_2 = ({x_2}^1, {x_2}^2,..., {x_2}^8)$ derived from a given scan, let us define its set of target labels $Y_2 = ({Y_2}^1, {Y_2}^2,..., {Y_2}^8) \in \{0,1\}$. Using a pair of complementary class weights to adjust for the patch-level class imbalance (1:4), we formulate the balanced cross-entropy loss function of $M_2$ for a single patch in each scan, as follows:

{\setlength{\abovedisplayskip}{-3pt}
\setlength{\belowdisplayskip}{1pt}
\begin{equation*}
\begin{split}
BCE({x_2}^i,{Y_2}^i) = &-\beta{Y_2}^i\text{log}{y_2}^i \\
                       &-(1-\beta)(1-{Y_2}^i)\text{log}(1-{y_2}^i) \indent i \in [1,8] 
\end{split}
\end{equation*}}

\noindent Here, ${y_2}^i = p(O\text{=}1|{x_2}^i) \in [0,1]$, represents the probability of ${x_2}^i$ being a \textit{malignant} patch as predicted by $M_2$. At test-time, $y_2 = ({y_2}^1, {y_2}^2,..., {y_2}^8) \in [0,1]$, i.e. a set of probabilistic malignancy scores for $x_2$, serves as the final output of $M_2$ for each scan.

Transforming voxel-level annotations into patch-wise labels can introduce additional noise in the target labels used at train-time. For instance, a single octant patch contains $64\times64\times8$ or 32768 voxels per channel. In a naive patch extraction system, if the fully delineated ground-truth for this sample includes even a single voxel of \textit{malignant} tissue, then the patch-wise label would be inaccurately set as \textit{malignant}, despite a voxel-level imbalance of 1:32767 supporting the alternate class. Such a training pair carries high label noise and proves detrimental to the learning cycle, where the network associates semantic features to the wrong target class. Therefore, we define a constraint $\tau$, representing the minimum percentage of \textit{malignant} tissue voxels required to assign the label \textit{malignant}.

For $M_2$, we consider CNN architectures based on residual learning for feature extraction, due to their modularity and continued success in supporting state-of-the-art segmentation and detection performance in the medical domain \citep{UoT_PCa_CLF,GoogleMammo,BraTS2019}.

\subsubsection{Decision Fusion}
\label{sec2.2.3}
The goal of the decision fusion node $N_{DF}$ is to aggregate $M_1$ and $M_2$ predictions ($y_1, y_2$) into a single output $y_{DF}$, which retains the same sensitivity as $y_1$, but improves specificity by reducing false positives. False positives in $y_1$ are fundamentally clusters of positive values located in the \textit{benign} regions of the scan. $N_{DF}$ employs $y_2$ as a means of identifying these regions. We set a threshold $T_P$ on $(1-{y_2}^i)$ to classify each patch ${x_2}^i$, where $i \in $[1,8]. $T_P$ represents the minimum probability required to classify ${x_2}^i$ as a \textit{benign} patch. A high value of $T_P$ adapts $M_2$ as a highly sensitive classifier that yields very few false negatives, if any at all. Once all \textit{benign} regions have been identified, any false positives within these patches are suppressed by multiplying their corresponding regions in $y_1$ with a penalty factor $\lambda$. Resultant detection map $y_{DF}$, i.e. essentially a post-processed $y_1$, serves as the final output of our proposed CAD system. Limiting $N_{DF}$ to a simple framework of two hyperparameters alleviates the risk of overfitting. While $T_P$ addresses which patches should be considered for false positive reduction, $\lambda$ addresses what factor candidate false positives should be suppressed by, within these selected patches. An appropriate combination of $T_P$ and $\lambda$ can either suppress clear false positives or facilitate an aggressive reduction scheme at the expense of fewer true positives in $y_{DF}$. In this research, we opted for the former policy to retain maximum csPCa detection sensitivity. Optimal values of $T_P$ and $\lambda$ were determined to be 0.98 and 0.90, respectively, via a coarse-to-fine hyperparameter grid search (detailed in the \textit{Supplementary Materials}).

\subsubsection{Anatomical Prior}
\label{sec2.2.4}
Parallel to recent studies in medical image computing \citep{DenseVNet, AnatomicalPriors, SpectralBrain, WeakProstate, PrioriAug} on infusing clinical priori into CNN architectures, we hypothesize that $M_1$ can benefit from an explicit anatomical prior for csPCa detection in bpMRI. To this end, we construct a probabilistic population prior $P \in [0,1]$, measuring $144\times144\times18$ voxels, as introduced in our previous work \citep{EncodClin}. $P$ captures the spatial prevalence and zonal distinction of csPCa using 1584 radiologically-estimated csPCa$^{(\textit{PR})}$ annotations and CNN-generated prostate zonal segmentations, respectively, from the training dataset. We opt for an early fusion technique to encode the priori \citep{ZonalPrior}, where $P$ is concatenated as an additional channel to every input scan passed through $M_1$, thereby guiding its learning cycle as a spatial weight map embedded with domain-specific clinical knowledge (refer to \hyperref[fig2]{Fig. 2}). Prior-to-image correspondence is established at both train-time and inference by using case-specific prostate segmentations to translate, orient and scale $P$ (via matching object centroids and maximizing volumetric overlap), accordingly, for each bpMRI scan.

%%% ----------------------------------------------

\subsection{Experimental Design}
\label{sec2.3}
Several experiments were conducted to statistically evaluate performance and analyze the design choices throughout the end-to-end model. We facilitated a fair comparison by maintaining an identical preprocessing, augmentation, tuning and train-validation pipeline for each candidate system in a given experiment. Patient-based diagnosis performance was evaluated using the Receiver Operating Characteristic (ROC), where the area under ROC (AUROC) was estimated from the normalized Wilcoxon/Mann-Whitney $U$ statistic \citep{U_stat}. Lesion-level performance was evaluated using the Free-Response Receiver Operating Characteristic (FROC) to address PCa multifocality. Detections sharing a minimum DSC with their ground-truth delineations were considered true positives. Given that the vast majority of csPCa lesions are small (median volume $<$ 2 cm$^3$ across both centers), have indistinct margins and share large inter-reader variability in their interpretation, we set this threshold as 0.10 DSC, in agreement with \cite{GoogleMammo}. All metrics were computed in 3D, across complete image volumes. Confidence intervals were estimated as twice the standard deviation from the mean of 5-fold cross-validation (applicable to validation sets) or  1000 replications of bootstrapping (applicable to testing sets) --as per standard practice \citep{UCLA,GoogleMammo}. Statistically significant improvements were verified with a $p$-value on the difference in case-level AUROC and lesion-level sensitivity at clinically relevant false positive rates (0.5, 1.0) using 1000 replications of bootstrapping \citep{pValue}. Bonferroni correction was used to adjust the significance level for multiple comparisons.

%% -------------------------------------------------
%% EXPERIMENTS & RESULTS ---------------------------
%% -------------------------------------------------
\section{Results and Analysis}
\label{sec3}
\subsection{Effect of Architecture and Label Noise on Classification}
\label{sec3.1}
To determine the effect of the classification architecture for $M_2$, five different 3D CNNs (ResNet-v2, Inception-ResNet-v2, Residual Attention Network, SEResNet, SEResNeXt) were implemented and tuned across their respective hyperparameters to maximize patient-based AUROC over 5-fold cross-validation. Furthermore, each candidate CNN was trained using whole-images and patches, in separate turns, to draw out a comparative analysis surrounding the merits of spatial context versus localized labels. In the latter case, we studied the effect of $\tau$ on patch-wise label assignment (refer to \hyperref[sec2.2.2]{Section 2.2.2}). We investigated four different values of $\tau$: 0.0\%, 0.1\%, 0.5\%, 1.0\%; which correspond to minimum csPCa volumes of 9, 297, 594 and 1188 mm$^3$, respectively. Each classifier was assessed qualitatively via 3D GradCAMs \citep{GradCAM} to ensure adequate interpretability for clinical usage.

\begin{figure*}[t!]
\centering
\includegraphics[width=1.0\textwidth]{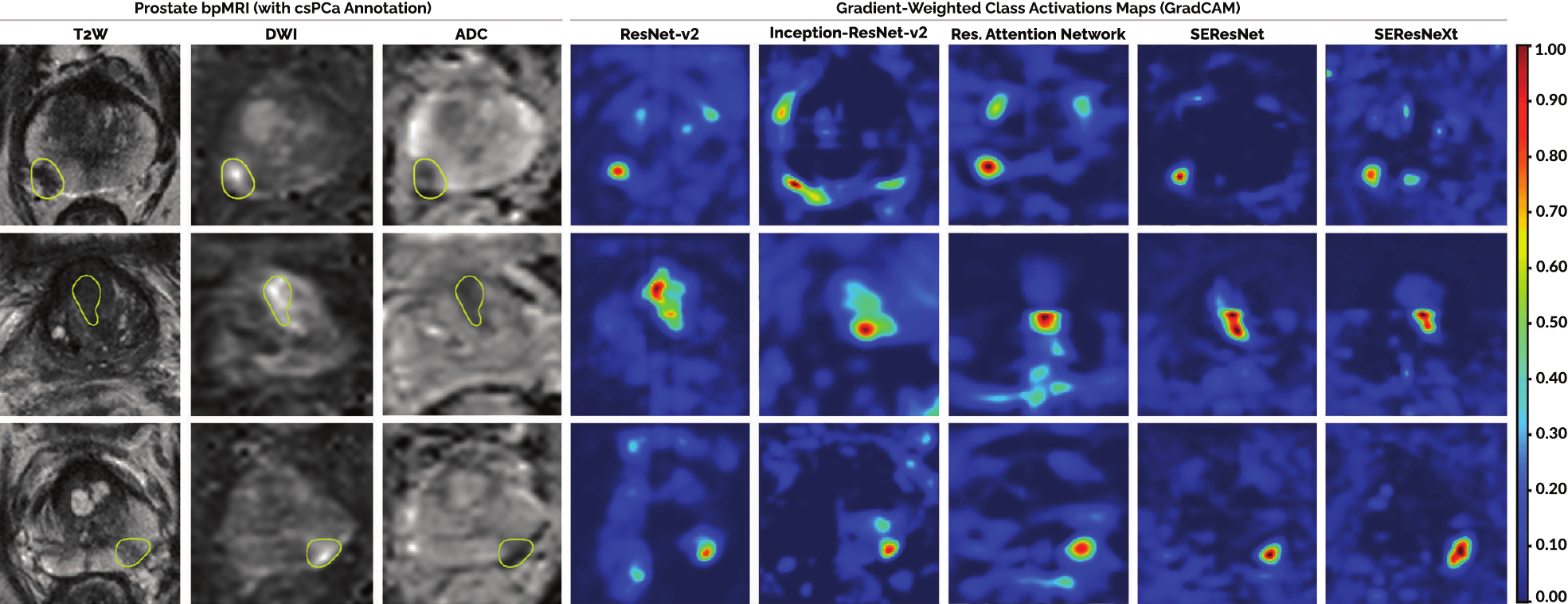}
\caption{Model interpretability of the candidate CNN architectures for classifier $M_2$ at $\tau=$0.1\%. Gradient-weighted class activation maps (GradCAM) and their corresponding T2W, DWI and ADC scans for three patient cases from the validation set are shown above. Each case included a single instance of csPCa$^{(\textit{PR})}$ located in the prostatic TZ \textbf{(\textit{center row})} or PZ \textbf{(\textit{top, bottom rows})}, as indicated by the yellow contours. Whole-image GradCAMs were generated by restitching and normalizing (\textit{min-max}) the eight patch-level GradCAMs generated per case. Maximum voxel-level activation was observed in close proximity of csPCa$^{(\textit{PR})}$, despite training each network using patch-level binary labels only.} 
\label{fig4}
\end{figure*}

From the results noted in \hyperref[tab1]{Table 1}, we observed that the SEResNet architecture consistently scored the highest AUROC across every training scheme. However, in each case, its performance remained statistically similar ($p \geq$ 0.01) to the other candidate models. We observed that a higher degree of supervision from patch-wise training proved more useful than the near 8$\times$ additional spatial context provided per sample during whole-image training. Increasing the value of $\tau$ consistently improved performance for all candidate classifiers (upto 10\% in patch-level AUROC). While we attribute this improvement to lower label noise at train-time, it is important to note that the total csPCa volume per patient is typically small (refer to \hyperref[sec2.1.2]{Section 2.1.2}). If $\tau$ is set too large, not only are patch labels regulated, as intended, but multiple patch-level label swaps can compound to the point where entire patient cases can swap labels --resulting in an inaccurate evaluation. Such a phenomenon was observed for 9 patient cases across the 1950 training-validation scans at $\tau=$\{0.5, 1.0\}\%. At $\tau=$0.1\%, there were no cases of patient-level label swaps (as seen at $\tau=$\{0.5, 1.0\}\%), while patch-level AUROC still improved by nearly 2\% relative to a naive patch extraction system ($\tau=$0.0\%). Thus, for the 3D CAD system, we chose the SEResNet patch-wise classifier trained at $\tau=$0.1\% as $M_2$. GradCAMs confirm that $M_2$ accurately targets csPCa$^{(\textit{PR})}$ lesions (if any) on a voxel-level basis, despite being trained on patch-level binary labels (as highlighted in \hyperref[fig4]{Fig. 4}). Further details regarding the network and training configurations of $M_2$ are listed in the \textit{Supplementary Materials}.

\begin{table*}[t!]
\renewcommand{\arraystretch}{1.00}
\caption{Patient-based diagnosis performance of the candidate CNN architectures and training schemes (whole-image versus patch-wise training with four different values of $\tau$ to regulate label noise) for classifier $M_2$. Performance scores indicate mean of 5-fold cross-validation, followed by 95\% confidence intervals estimated as twice the standard deviation.}
\vspace{2mm}
{\small

\begin{tabular}{
p{0.335\textwidth}>
{\centering}p{0.0595\textwidth}>
{\centering}p{0.120\textwidth}>
{\centering}p{0.0815\textwidth}>
{\centering}p{0.0815\textwidth}>
{\centering}p{0.0815\textwidth}>
{\centering\arraybackslash}p{0.081\textwidth}}

\toprule
\multirow{2}{*}{Model}    &
\multirow{2}{*}{Params}   &
\multirow{2}{*}{AUROC}    &
\multicolumn{4}{c}{AUROC (Patches)} \\

\cline{4-7}
&
& {(Whole-Image)}
& \rule{0pt}{11.5pt} $\tau=0.0$\%  
& \rule{0pt}{11.5pt} $\tau=0.1$\%  
& \rule{0pt}{11.5pt} $\tau=0.5$\%  
& \rule{0pt}{11.5pt} $\tau=1.0$\% \\

\midrule
ResNet-v2 \citep{ResNetv2}                   & 0.089 M & $0.819 {\scriptstyle \pm 0.018}$ & $0.830 {\scriptstyle \pm 0.010}$ & $0.844 {\scriptstyle \pm 0.011}$ & $0.868 {\scriptstyle \pm 0.013}$ & $0.897 {\scriptstyle \pm 0.008}$\\
Inception-ResNet-v2 \citep{Inception-ResNet} & 6.121 M & $0.823 {\scriptstyle \pm 0.017}$ & $0.822 {\scriptstyle \pm 0.014}$ & $0.860 {\scriptstyle \pm 0.015}$ & $0.883 {\scriptstyle \pm 0.009}$ & $0.905 {\scriptstyle \pm 0.008}$\\
Res. Attention Network \citep{ResAtnNet}     & 1.233 M & $0.826 {\scriptstyle \pm 0.024}$ & $0.837 {\scriptstyle \pm 0.012}$ & $0.850 {\scriptstyle \pm 0.007}$ & $0.876 {\scriptstyle \pm 0.008}$ & $0.901 {\scriptstyle \pm 0.008}$\\
SEResNet \citep{SE}                          & 0.095 M & $0.836 {\scriptstyle \pm 0.014}$ & $0.842 {\scriptstyle \pm 0.019}$ & $0.861 {\scriptstyle \pm 0.005}$ & $0.886 {\scriptstyle \pm 0.008}$ & $0.912 {\scriptstyle \pm 0.008}$\\
SEResNeXt \citep{SE}                         & 0.128 M & $0.820 {\scriptstyle \pm 0.022}$ & $0.833 {\scriptstyle \pm 0.013}$ & $0.843 {\scriptstyle \pm 0.005}$ & $0.875 {\scriptstyle \pm 0.009}$ & $0.896 {\scriptstyle \pm 0.012}$\\
\bottomrule
\end{tabular}
}
\label{tab1}
\end{table*}

\subsection{Effect of Architecture and Clinical Priori on Detection}
We analyzed the effect of the $M_1$ architecture, in comparison to the four baseline 3D CNNs (U-SEResNet, UNet++, nnU-Net, Attention U-Net) that inspire its design. We evaluated the end-to-end 3D CAD system, along with the individual contributions of its constituent components ($M_1$, $M_2$, $P$), to examine the effects of false positive reduction and clinical priori. Additionally, we applied the ensembling heuristic of the nnU-Net framework \citep{nnUNet} to create CAD$\Conv$, i.e. an ensemble model comprising of multiple CAD instances, and we studied its impact on overall performance. Each candidate setup was tuned over 5-fold cross-validation and benchmarked on the testing datasets (TS1, TS2).

\begin{figure*}[!t]
\centering

\includegraphics[width=\textwidth]{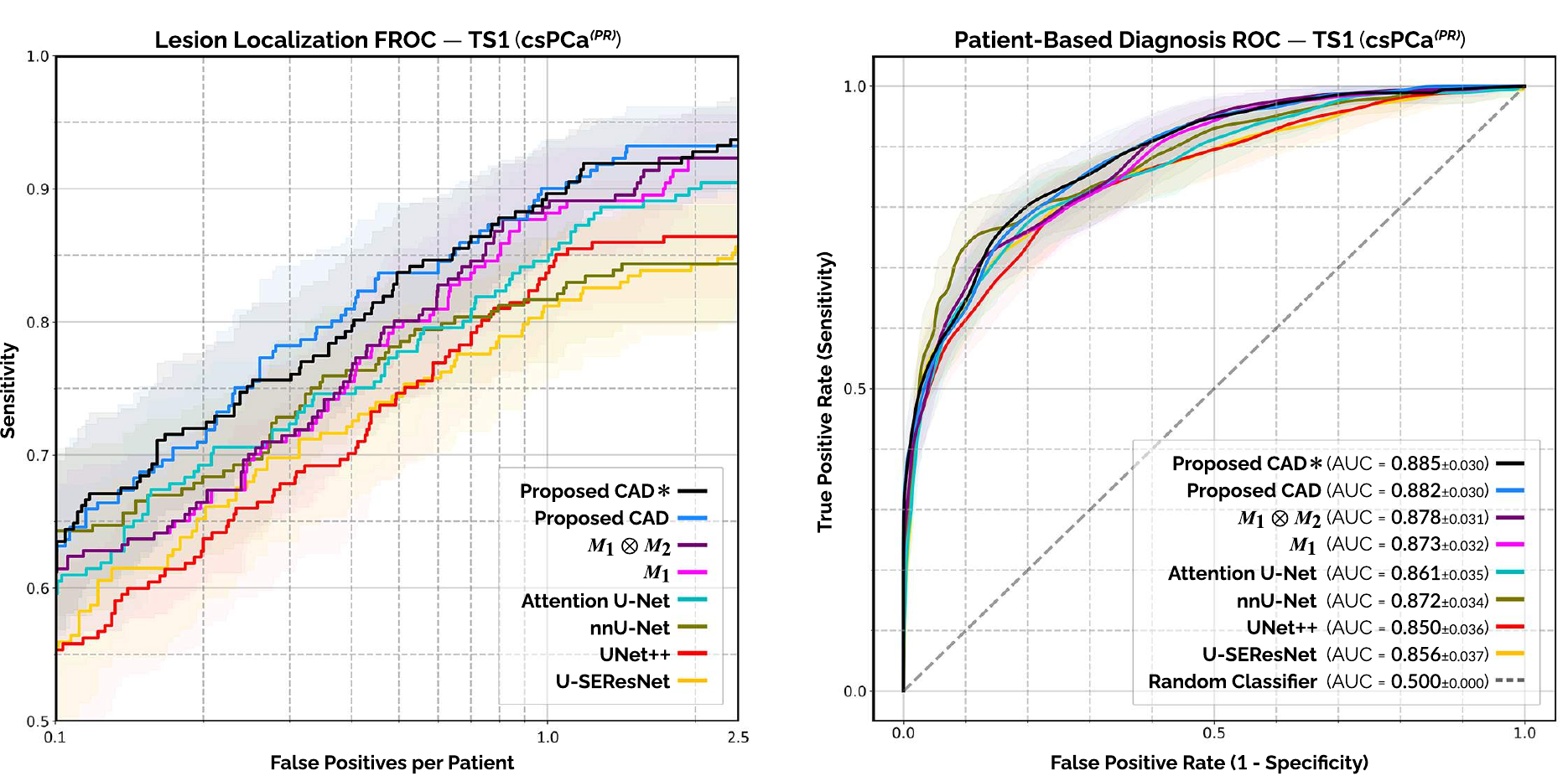}

\vspace{4mm}

\includegraphics[width=\textwidth]{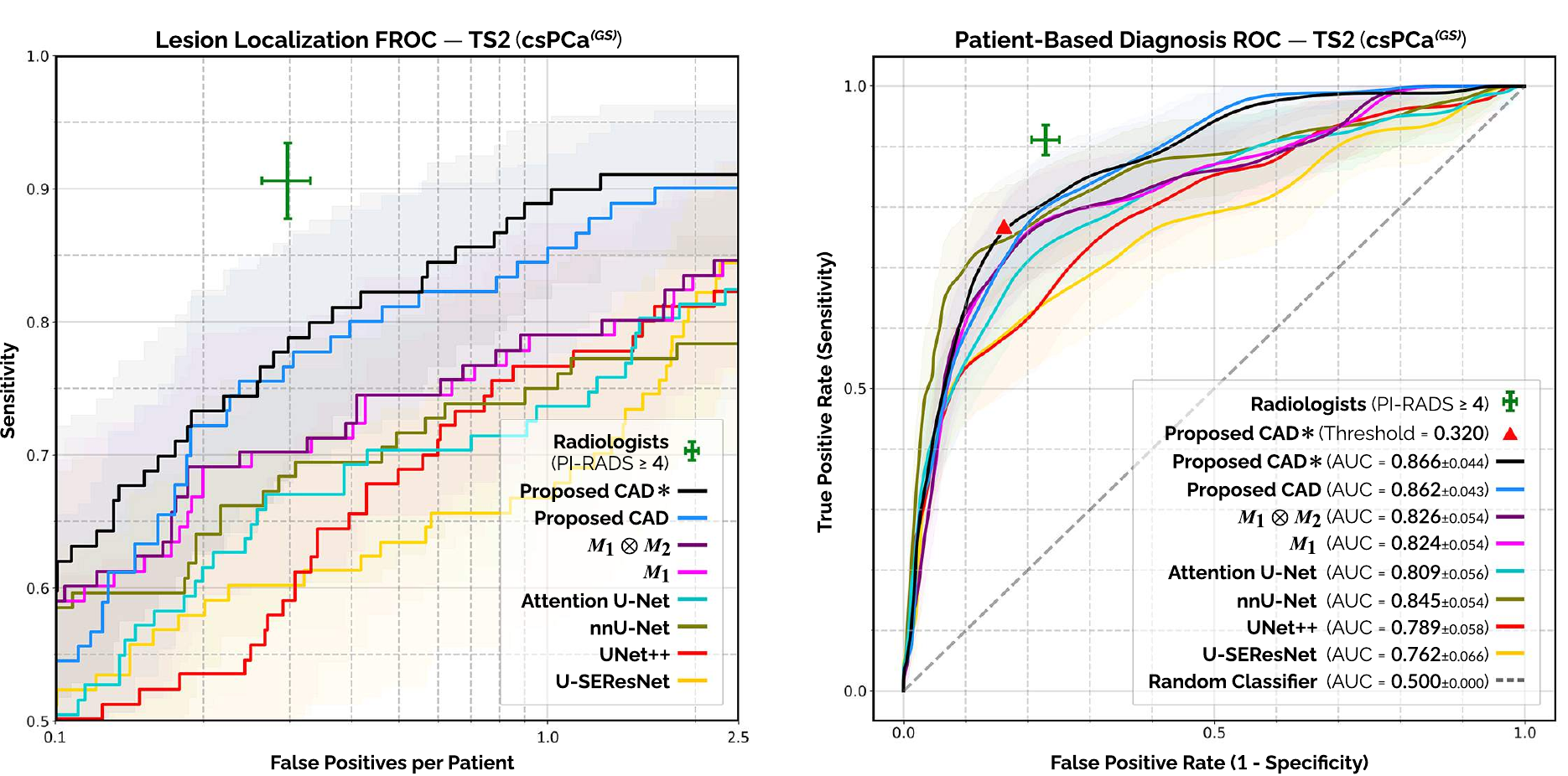}

\caption{Lesion-level FROC \textbf{(\textit{left})} and patient-based ROC \textbf{(\textit{right})} analyses of csPCa$^{(\textit{PR})}$ \textbf{(\textit{top row})} / csPCa$^{(\textit{GS})}$ \textbf{(\textit{bottom row})} detection sensitivity against the number of false positives generated per patient scan using the baseline, ablated and proposed detection models on the institutional testing set TS1 \textbf{(\textit{top row})} and the external testing set TS2 \textbf{(\textit{bottom row})}. Transparent areas indicate the 95\% confidence intervals. Mean performance for the consensus of expert radiologists and their 95\% confidence intervals are indicated by the centerpoint and length of the green markers, respectively, where all observations marked PI-RADS 4 or 5 are considered positive detections (as detailed in \hyperref[sec2.3]{Section 2.3}).}
\label{fig5}
\end{figure*}

\subsubsection{Generalization to Radiologically-Estimated csPCa}
\label{sec3.2.1}
\begin{figure}[!t]
\centering
\includegraphics[width=0.478\textwidth]{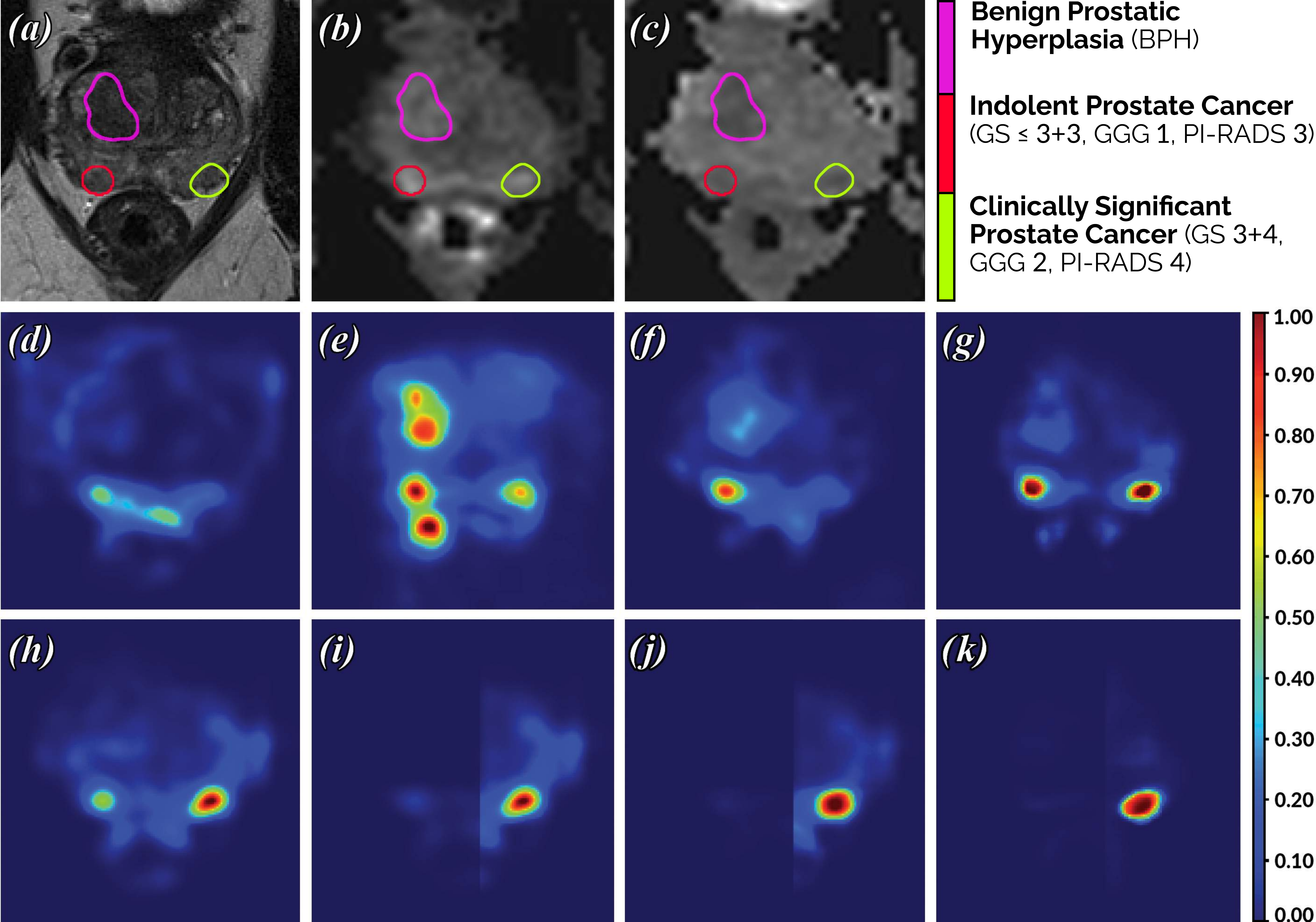}
\caption{\textbf{(\textit{a})} T2W, \textbf{(\textit{b})} DWI, \textbf{(\textit{c})} ADC scans for a patient case in the external testing set TS2, followed by its csPCa detection map as predicted by each candidate system: \textbf{(\textit{d})} U-SEResNet, \textbf{(\textit{e})} UNet++, \textbf{(\textit{f})} Attention U-Net, \textbf{(\textit{g})} nnU-Net, \textbf{(\textit{h})} $M_1$, \textbf{(\textit{i})} $M_1 \otimes M_2$, \textbf{(\textit{j})} proposed CAD, \textbf{(\textit{k})} proposed CAD$\protect\Conv$. Three stand-alone detection networks (UNet++, nnU-Net, $M_1$) successfully identified the csPCa$^{(\textit{GS})}$ lesion, albeit with additional false positive(s). In the case of the proposed CAD/CAD$\protect\Conv$ system, while the classifier in $M_1 \otimes M_2$ was able to suppresses these false positive(s) from $M_1$, inclusion of prior $P$ further strengthened the confidence and boundaries of the true positive. At the same time, it should be noted that small lesions carry higher uncertainty in their grading --i.e. indolent GGG 1 lesions may create suspicious findings in MRI, while clinically significant GGG 2 lesions risk being undersampled during biopsy \citep{BiopsySOTA,Overdiag}.}
\vspace{-4mm}
\label{fig6}
\end{figure}

\noindent \textbf{Lesion Localization}: From the FROC analysis on the institutional testing set TS1 (refer to \hyperref[fig5]{Fig. 5}), we observed that $M_1$ reached 88.15$\pm$4.19\% detection sensitivity at 1.0 false positive per patient, significantly ($p$ $\leq$ 0.01) outperforming the baseline U-SEResNet (81.18$\pm$4.99\%), UNet++ (83.81$\pm$4.80\%), nnU-Net (81.67$\pm$4.64\%) and Attention U-Net (84.76$\pm$4.64\%). With the addition of classifier $M_2$ to $M_1$ ($M_1 \otimes M_2$), upto 12.89\% ($p$ $\leq$ 0.001) less false positives were generated per patient, while retaining the same maximum detection sensitivity (92.29\%) as before. The working principle of $M_1 \otimes M_2$ is illustrated in \hyperref[fig6]{Fig. 6} through a particularly challenging patient case, where the prostate gland is afflicted by multiple, simultaneously occurring conditions. With the inclusion of anatomical prior $P$ in $M_1 \otimes M_2$, the 3D CAD system benefited from a further 3.14\% increase in partial area under FROC (pAUC) between 0.10--2.50 false positives per patient, reaching 1.676$\pm$0.078 pAUC. At 0.5 false positive per patient, the 3D CAD system reached 83.69$\pm$5.22\% detection sensitivity, surpassing the best baseline (nnU-Net) by 5.59\% ($p$ $\leq$ 0.001), while detecting 4.10\% ($p$ $\leq$ 0.01) and 3.63\% ($p$ $\leq$ 0.01) more csPCa$^{(\textit{PR})}$ lesions than its component systems $M_1$ and $M_1 \otimes M_2$, respectively. It reached a maximum detection sensitivity of 93.19$\pm$2.96\% at 1.46 false positives per patient, identifying a higher percentage of csPCa occurrences than all other candidate systems.

\noindent \textbf{Patient-Based Diagnosis}: From ROC analysis on the institutional testing set TS1 (refer to \hyperref[fig5]{Fig. 5}), we observed that the 3D CAD system reached 0.882$\pm$0.03 AUROC in case-level diagnosis, ahead of all other candidate systems by a margin of 0.4--3.2\%. While it performed significantly better than the baseline U-SEResNet ($p$ $\leq$ 0.01), UNet++ ($p$ $\leq$ 0.001) and Attention U-Net ($p$ $\leq$ 0.01), its ability to discriminate between \textit{benign} and \textit{malignant} patient cases was statistically similar ($p$ $\geq$ 0.01) to the nnU-Net, $M_1$ and $M_1 \otimes M_2$.

\begin{table*}[!b]
\renewcommand{\arraystretch}{1.00}
\caption{Computational requirements (in terms of the number of trainable parameters, VRAM usage and the average time taken per patient scan during inference on a single NVIDIA RTX 2080 Ti) against the localization performance (in terms of the maximum csPCa detection sensitivity achieved and its corresponding false positive rate across both testing datasets) for each candidate detection system.}
\vspace{2mm}
{\small

\begin{tabular}{
p{0.334\textwidth}>
{\raggedleft}p{0.0805\textwidth}>
{\centering}p{0.0650\textwidth}>
{\centering}p{0.0775\textwidth}>
{\centering}p{0.1420\textwidth}>
{\centering \arraybackslash}p{0.1420\textwidth}}

\toprule
\multirow{2}{*}{Model}                                           & 
\multirow{2}{*}{Params \hspace{0.75pt}}                          & 
\multirow{2}{*}{VRAM}                                            & 
\multirow{2}{*}{Inference}                                       &
\multicolumn{2}{c}{Maximum Sensitivity \{False Positive Rate\}}  \\
\cline{5-6}
&
& 
&
& \rule{0pt}{11.5pt} TS1 -- csPCa$^{(\textit{PR})}$  
& \rule{0pt}{11.5pt} TS2 -- csPCa$^{(\textit{GS})}$\\

\midrule
U-SEResNet       \citep{SE}                                        &  1.615 M  & $0.94$ GB & $1.77\scriptstyle \pm 0.20$ s  & $85.63\% {\scriptstyle \pm 4.70}$ $\{2.44\}$ & $84.42\% {\scriptstyle \pm 7.36}$ $\{2.26\}$\\
UNet++           \citep{UNet++}                                    & 14.933 M  & $2.97$ GB & $1.79\scriptstyle \pm 0.19$ s  & $86.41\% {\scriptstyle \pm 4.54}$ $\{1.74\}$ & $82.28\% {\scriptstyle \pm 7.62}$ $\{2.25\}$\\
nnU-Net          \citep{nnUNet}                                    & 30.599 M  & $4.69$ GB & $2.09\scriptstyle \pm 0.03$ s  & $84.34\% {\scriptstyle \pm 4.40}$ $\{1.44\}$ & $77.23\% {\scriptstyle \pm 8.14}$ $\{1.12\}$\\
Attention U-Net  \citep{GridAtn-Gates}                             &  2.235 M  & $1.96$ GB & $1.77\scriptstyle \pm 0.19$ s  & $90.46\% {\scriptstyle \pm 3.63}$ $\{2.07\}$ & $82.43\% {\scriptstyle \pm 7.79}$ $\{2.32\}$\\
\hline
Dual-Attention U-Net -- $M_1$                                      & 15.250 M  & $3.01$ GB & $1.79\scriptstyle \pm 0.19$ s  & $92.29\% {\scriptstyle \pm 3.24}$ $\{1.94\}$ & $84.60\% {\scriptstyle \pm 7.45}$ $\{2.31\}$\\
$M_1$ with \textit{False Positive Reduction} -- $M_1 \otimes M_2$  & 15.335 M  & $3.75$ GB & $1.89\scriptstyle \pm 0.23$ s  & $92.29\% {\scriptstyle \pm 3.24}$ $\{1.69\}$ & $84.60\% {\scriptstyle \pm 7.45}$ $\{2.22\}$\\
$M_1 \otimes M_2$ with \textit{Prior} -- Proposed CAD              & 15.335 M  & $3.98$ GB & $1.90\scriptstyle \pm 0.23$ s  & $93.19\% {\scriptstyle \pm 2.96}$ $\{1.46\}$ & $90.03\% {\scriptstyle \pm 5.80}$ $\{1.67\}$\\
Ensemble of CAD -- Proposed CAD$\Conv$                             & 40.069 M  & $9.85$ GB & $2.41\scriptstyle \pm 0.42$ s  & $93.69\% {\scriptstyle \pm 3.13}$ $\{2.36\}$ & $91.05\% {\scriptstyle \pm 5.24}$ $\{1.29\}$\\
\bottomrule
\end{tabular}
}
\label{tab2}
\end{table*}

\subsubsection{Generalization to Histologically-Confirmed csPCa}
\label{sec3.2.2}
Both the FROC and ROC analyses on the external testing set TS2 (refer to \hyperref[fig5]{Fig. 5}) indicate similar patterns emerging as those observed in \hyperref[sec3.2.1]{Section 3.2.1}, but with an overall decrease in performance. Given the near-identical MRI scanners and acquisition conditions employed between both institutions (refer to \hyperref[sec2.1.1]{Section 2.1.1}), we primarily attribute this decline to the disparity between the imperfect radiologically-estimated training annotations (csPCa$^{(\textit{PR})}$) and the histologically-confirmed testing annotations (csPCa$^{(\textit{GS})}$) in TS2 (refer to \hyperref[sec3.3]{Section 3.3} for radiologists' performance). By comparing the relative drop in performance for each candidate model, we can effectively estimate their generalization and latent understanding of csPCa, beyond our provided training samples.   

\vspace{1mm}

\noindent \textbf{Lesion Localization}: At 1.0 false positive per patient, the 3D CAD system achieved 85.55$\pm$7.04\% detection sensitivity on TS2 (refer to \hyperref[fig5]{Fig. 5}), performing significantly better ($p$ $\leq$ 0.001) than the baseline U-SEResNet (66.74$\pm$9.65\%), UNet++ (76.66$\pm$9.05\%), nnU-Net (74.73$\pm$7.72\%) and Attention U-Net (73.64$\pm$8.97\%). It also detected 6.56\% ($p$ $\leq$ 0.005) more csPCa$^{(\textit{GS})}$ lesions than its ablated counterparts $M_1$ and $M_1 \otimes M_2$, respectively. The 3D CAD system reached a maximum detection sensitivity of 90.03$\pm$5.80\% at 1.67 false positives per patient, scoring higher than all other candidate systems. On average, all baseline models underwent 7--13\% drops in detection sensitivity at 1.0 false positive per patient, relative to their performance on TS1. Similarly, the average detection sensitivities of $M_1$ and $M_1 \otimes M_2$ fell by nearly 10\%. From the inclusion of $P$ in $M_1 \otimes M_2$, this decline came down to only 3\% for the 3D CAD system at the same false positive rate. Furthermore, an overall 11.54\% increase in pAUC was observed between 0.10--2.50 false positives per patient, relative to $M_1 \otimes M_2$. 

\vspace{1mm}

\noindent \textbf{Patient-Based Diagnosis}: 3D CAD reached 0.862$\pm$0.04 AUROC on TS2 (refer to \hyperref[fig5]{Fig. 5}), ahead of the baseline U-SEResNet, UNet++, nnU-Net and Attention U-Net by 10.0\% ($p \leq$ 0.001), 7.3\% ($p \leq$ 0.001), 1.7\% ($p$ $>$ 0.1) and 5.3\% ($p$ $\leq$ 0.05), respectively. Compared to TS1, the 3D CAD model underwent 2\% decrease in AUROC, while all other candidate systems underwent an average reduction of 5--6\%. Once again, the anatomical prior proved vital, enabling 3D CAD to outperform its immediate counterpart $M_1 \otimes M_2$ by 3.6\% ($p$ $\leq$ 0.05).

\subsubsection{Effect of Ensembling}
\label{sec3.2.3}
The ensembled prediction of CAD$\Conv$ is the weighted-average output of three member models: 2D, 3D and two-stage cascaded 3D variants of the proposed CAD system (refer to \textit{Supplementary Materials} for detailed implementation). In comparison to the standard 3D CAD system, CAD$\Conv$ carries 2.6$\times$ trainable parameters, occupies 2.5$\times$ VRAM for hardware acceleration and requires 1.3$\times$ inference time per patient scan (as noted in \hyperref[tab2]{Table 2}). In terms of its performance, CAD$\Conv$ demonstrated 0.3--0.4\% improvement in patient-based AUROC across both testing datasets and shared statistically similar lesion localization on TS1. It generated a considerably large improvement in lesion detection on TS2, amounting to 4.01\% increase in pAUC between 0.10--2.50 false positives per patient (refer to \hyperref[fig5]{Fig 5}), as well as a higher maximum detection sensitivity (91.05$\pm$5.24\%) at a lower false positive rate (1.29) (as noted in \hyperref[tab2]{Table 2}).

\subsection{Relative Performance to Consensus of Radiologists}
\label{sec3.3}
To evaluate CAD$\Conv$ in comparison to the consensus of expert radiologists, we analyzed their relative performance on the external testing set TS2. Agreements in patient-based diagnosis were computed with Cohen's \textit{kappa}.

\begin{figure*}[!t]
\centering

\includegraphics[width=0.85\textwidth]{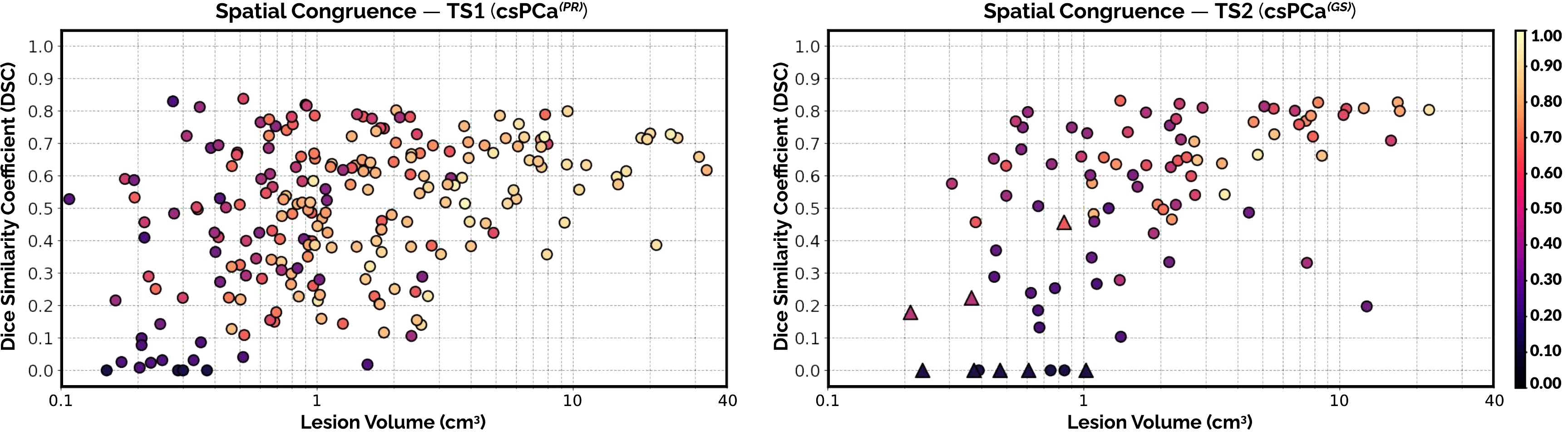}

\caption{Distribution of per-lesion Dice Similarity Coefficient (DSC) (relative to csPCa lesion volume) for CAD$\Conv$ detections against the ground-truth annotations of csPCa$^{(\textit{PR})}$ in the institutional testing TS1 \textit{\textbf{(left)}} and csPCa$^{(\textit{GS})}$ in the external testing set TS2 \textit{\textbf{(right)}}. All DSC values were computed in 3D for the model-specific operating point with maximum detection sensitivity (91.05$\pm$5.24$\%$). Encoded color for each marker indicates its corresponding likelihood of malignancy, as predicted by CAD$\Conv$. Triangular markers for TS2 \textit{\textbf{(right)}} indicate csPCa$^{(\textit{GS})}$ lesions missed by the radiologists.}
\label{fig7}
\end{figure*}

Radiologists achieved 90.72$\pm$2.78\% detection sensitivity at 0.30 false positives per patient and 91.11$\pm$2.67\% sensitivity at 77.18$\pm$2.37\% specificity in lesion localization and patient-based diagnosis, respectively (refer to \hyperref[fig5]{Fig. 5}). After binarizing its case-level detections using a probability threshold ($>$0.32) \citep[as demonstrated by][]{RadPRvsDL}, the CAD$\Conv$ system reached 75.31$\pm$3.64\% sensitivity at 85.83$\pm$2.22\% specificity in patient-based diagnosis, where it shared 76.69\% (227/296 cases; \textit{kappa} $=$ $0.51\pm0.04$) and 81.08\% (240/296 cases; \textit{kappa} $=$ $0.56\pm0.06$) agreement with expert radiologists and independent pathologists, respectively. In comparison, radiologists shared 81.42\% (241/296 cases; \textit{kappa} $=$ 0.61$\pm$0.05) agreement with pathologists. Pathologists often struggle to accurately differentiate GGG 1 from GGG 2 patterns in tissue specimens, resulting in $\geq30\%$ inter-reader variability \citep{GGG12_1,GGG12_2}. As observed in TS2, this challenge persists across prostate MRI, where radiologists' performance dropped to 88.24$\pm$2.67\% sensitivity and 51.67$\pm$3.45\% specificity, while discriminating patients with GGG 1 lesions (64/296 cases) from those carrying GGG 2 lesions (42/296 cases), exclusively. At the same specificity, CAD$\Conv$ performed statistically similar ($p$ $\geq$ 0.05), reaching 81.76$\pm$10.72\% sensitivity.

\subsection{Spatial Congruence Analysis}
\label{sec3.4}
On average, CAD$\Conv$ shared $0.49\pm0.22$ DSC and $0.58\pm0.21$ DSC on TS1 (csPCa$^{(\textit{PR})}$) and TS2 (csPCa$^{(\textit{GS})}$), respectively (refer to \hyperref[fig7]{Fig. 7}). In comparison, similar studies on csPCa$^{(\textit{GS})}$ detection from MRI reported 0.35--0.46 DSC \citep{DSC1, DSC2, DSC3, RadPRvsDL, DSC4}. Increasing model confidence and higher DSC values were observed consistently for larger lesions. However, DSC remains limited as an evaluation metric for detection --demonstrating high variance with minute changes in predicted contours and/or volume for smaller lesions \citep{DSC_LIM}. High DSC variance among smaller lesions could also be indicative of misaligned MRI sequences (potentially due to physical distortions from the low bandwidth of DWI) \citep{DWI-ART}. Unlike hard, binary masks; training CNNs with probabilistic target labels and output detections could potentially capture such uncertainty along lesion contours \citep{SoftSeg}.

\begin{figure*}[!b]
\centering
\includegraphics[width=\textwidth]{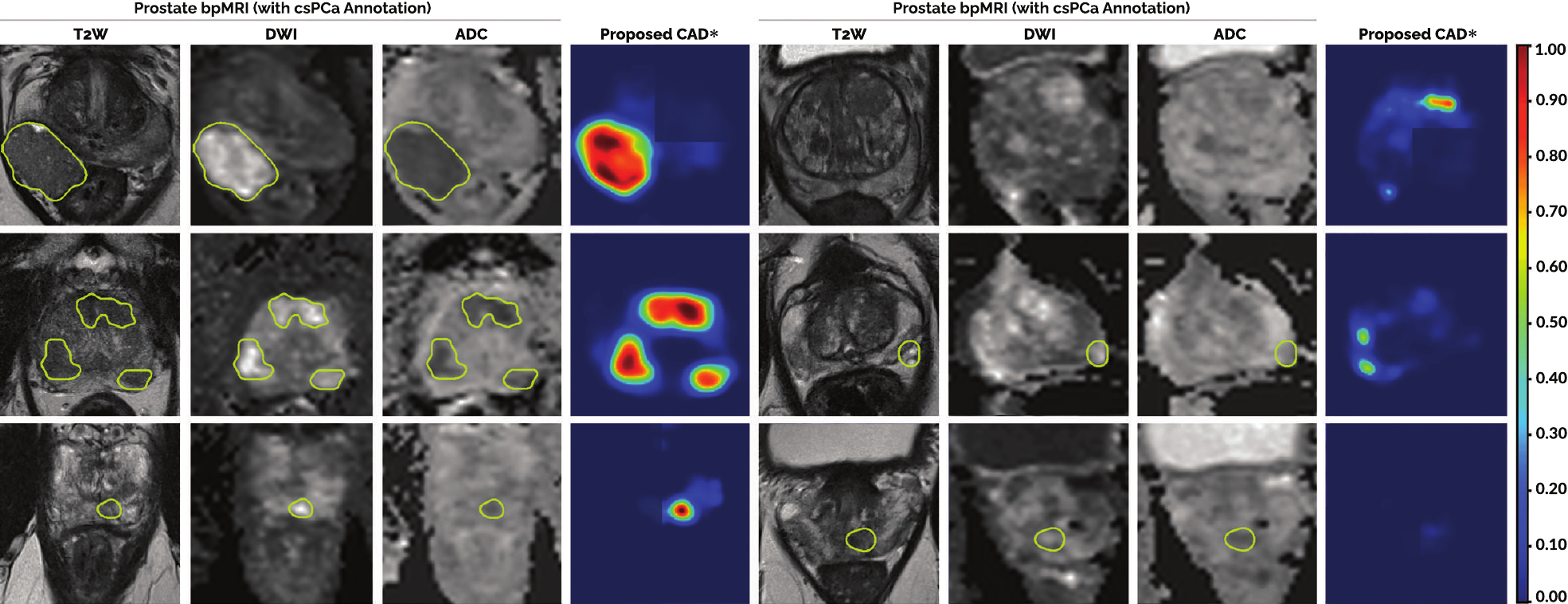}
\caption{Six patient cases from the external testing set TS2 and their corresponding csPCa detection maps, as predicted by the proposed CAD$\protect\Conv$ system. Yellow contours indicate csPCa$^{(\textit{GS})}$ lesions, if present. While CAD$\protect\Conv$ was able to successfully localize large, multifocal and apical/basal instances of csPCa$^{(\textit{GS})}$ \textbf{\textit{(left)}}, in the presence of severe inflammation/fibrosis induced by other non-malignant conditions (eg. BPH, prostatitis), CAD$\protect\Conv$ misidentified smaller lesions, resulting in false positive/negative predictions \textbf{\textit{(right)}}.}
\label{fig8}
\end{figure*}

%% -------------------------------------------------
%% DISCUSSION & CONCLUSION -------------------------
%% -------------------------------------------------

\section{Discussion and Conclusion}
\label{sec4}
We conclude that a detection network ($M_1$), harmonizing state-of-the-art attention mechanisms, can accurately discriminate more malignancies at the same false positive rate, than one without (refer to \hyperref[sec3.2.1]{Section 3.2.1}). Among four other recent adaptations of the 3D U-Net that are popularly used for biomedical segmentation, $M_1$ detected significantly more csPCa lesions at 1.00 false positive per patient and consistently reached the highest detection sensitivity on the testing datasets between 0.10--2.50 false positives per patient, closely followed by the Attention U-Net (refer to \hyperref[fig5]{Fig. 5}). As soft attention mechanisms continue to evolve, supporting ease of optimization, sharing equivariance over permutations \citep{InductiveBiases} and suppressing gradient updates from inaccurate annotations \citep{ResAtnNet, AAAI-Attention}, deep attentive models, such as $M_1$, become increasingly more applicable for csPCa detection in bpMRI \citep{SelfAttnPZ, Panoptic}. 

We conclude that a residual patch-wise 3D classifier ($M_2$) can significantly reduce false positives, without sacrificing high sensitivity. In stark contrast to ensembling, which scaled up the number of trainable parameters nearly 3$\times$ for limited improvements in performance (refer to \hyperref[sec3.2.3]{Section 3.2.3}), $M_2$ produced flat increases in specificity (upto 12.89\% less false positives per patient) across both testing datasets, while requiring less than 1\% of the total parameters in the 3D CAD system (as noted in \hyperref[tab2]{Table 2}). Furthermore, as a decoupled classifier, $M_2$ shares two major advantages. Firstly, unlike the jointly-trained, cascaded approach proposed by \cite{FPSiemens}, where the second-stage classifier was able to reduce false positives at the expense of nearly an 8\% decrease in detection sensitivity, in our case, the effect of $M_2$ on the overall 3D CAD system could be controlled via the decision fusion node $N_{DF}$, such that the maximum detection sensitivity of the system was completely retained (refer to \hyperref[tab2]{Table 2}). Secondly, due to its independent training scheme, $M_2$ remains highly modular, i.e. it can be easily tuned, upgraded or swapped out entirely upon future advancements, without retraining or affecting the stand-alone performance of $M_1$.

We conclude that encoding an anatomical prior ($P$) into the CNN architecture can guide model generalization with domain-specific clinical knowledge. Results indicated that $P$ played the most important role in the generalization of the 3D CAD system (via $M_1$) and in retaining its performance across the multi-institutional testing datasets (refer to \hyperref[sec3.2.2]{Section 3.2.2}). Remarkably, its contribution was substantially more than any other architectural enhancement proposed in recent literature, while introducing negligible changes in the number of trainable parameters (refer to \hyperref[tab2]{Table 2}). However, it is worth noting that similar experiments with classifier $M_2$, yielded no statistical improvements. Parallel to the methods proposed by \cite{RCNN} and \cite{NoduleNet}, $M_2$ was designed to learn a different set of feature representations for csPCa than $M_1$, using its smaller receptive field size, patch-wise approach and decoupled optimization strategy. Thus, while $M_1$ was trained to learn translation covariant features for localization, $M_2$ was trained to learn translation invariant features for classification, i.e. patch-wise prediction of the presence/absence of csPCa, irregardless of its spatial context in the prostate gland. We presume this key difference to be the primary reason why $M_2$ was effective at independent false positive reduction, yet unable to leverage the spatial priori embedded in $P$. Nonetheless, our study confirmed that powerful anatomical priors, such as $P$, can substitute additional training data for deep learning-based CAD systems and improve model generalization, by relaying the inductive biases of csPCa in bpMRI \citep{InductiveBiases}.

We hypothesized that modern CNN models, such as the CAD system presented in this study, should be able to learn and generalize beyond their training annotations of csPCa$^{(\textit{PR})}$ to reliably detect csPCa$^{(\textit{GS})}$ in an independent dataset \citep{NotSoSuper}. To test this hypothesis, we acquired a large cohort of 1950 patient scans and produced voxel-level delineations of all PI-RADS$>$3 lesions. In practice, as every prostate bpMRI scan shares a corresponding radiology report compliant with PI-RADS, our training cohort was able to effectively capture a broad range of highly diverse studies --representing the complete distribution of patients encountered in the clinical workflow, and not only those who underwent biopsies. We benchmarked the CAD system against a consensus of radiologists on the ZGT cohort --an external testing set (TS2) graded by independent pathologists. Radiologists operated with high sensitivity (refer to \hyperref[sec3.3]{Section 3.3}), detecting 216 PI-RADS $\geq 3$ lesion(s) among 152/296 patients (as noted in the \textit{Supplementary Materials}), who subsequently underwent both TRUS and MRI-guided biopsies. When jointly performed, these techniques have been reported to reach upto 97\% sensitivity for csPCa$^{(\textit{GS})}$ detection, relative to radical prostatectomy specimens \citep{TRUSMRGB_RP}. While cohorts of radical prostatectomy patients share potentially stronger ground-truth, they typically carry much fewer samples \citep{MultiInst, UCLA, CorrSig, SPCNet2021} and mostly, if not only, cases with severe malignancies --thereby deviating from the general distribution of patients encountered during clinical routine. Hence, an independent analysis of prostatectomy patients, exclusively, was not performed in this study \citep[in agreement with][]{RadPRvsDL}. For the ZGT cohort, we observed that CAD$\Conv$ demonstrated higher agreement with pathologists (81.08\%; \textit{kappa} $=$ $0.56\pm0.06$) than it did with radiologists (76.69\%; \textit{kappa} $=$ $0.51\pm0.04$). Moreover, CAD$\Conv$ detected csPCa$^{(\textit{GS})}$ lesions in three patient cases, which were missed by four experienced radiologists operating at high sensitivity. Notably, this verified that CNNs can train effectively on csPCa$^{(\textit{PR})}$ annotations and achieve competitive test performance for csPCa$^{(\textit{GS})}$ detection, in comparison to state-of-the-art solutions that exclusively utilize biopsy-confirmed training annotations \citep{MultiInst, UCLA, RadPRvsDL, UoT_PCa_CLF, TwoStep, CorrSig, Patch3D2020, SPCNet2021} (refer to \hyperref[sec1.1]{Section 1.1}). Although, CNNs remain inadequate as stand-alone solutions (refer to \hyperref[fig5]{Fig. 5}, \hyperref[fig8]{8}), the moderate agreement of CAD$\Conv$ with both clinical experts, while inferring predictions relatively dissimilar to radiologists, highlights its potential to improve diagnostic certainty as a viable second reader in a screening setting \citep{SanfordDLPR, ClinicalDep}.

The study is limited in a few aspects. Within the scope of this research, all prostate scans were acquired using MRI scanners developed by the same vendor. Thus, generalizing our proposed solution to a vendor-neutral model requires special measures, such as domain adaptation \citep{DomAdapProst}, to account for heterogeneous acquisition conditions. Inclusion of PI-RADS$>$3 lesions for constructing the anatomical prior may introduce bias at train-time, particularly if the corresponding radiologist(s) lack(s) substantial expertise. Radiologists utilize additional clinical variables (e.g. prior exams, DCE scans, PSA density levels, etc.) to inform their diagnosis for each patient case --limiting the equity of any direct comparisons against the 3D CNNs developed in this study.

In summary, an automated novel end-to-end 3D CAD system, harmonizing several state-of-the-art methods from recent literature, was developed to diagnose and localize csPCa in bpMRI. To the best of our knowledge, this was the first demonstration of a deep learning-based 3D detection and diagnosis system for csPCa trained using radiologically-estimated annotations only and evaluated on large, multi-institutional testing datasets. The promising results of this research motivate the ongoing development of new techniques, particularly those which factor in the breadth of clinical knowledge established in the field beyond limited training datasets, to create comprehensive CAD solutions for the clinical workflow of prostate cancer management.

\section*{Acknowledgements}
The authors would like to acknowledge Maarten de Rooij and Ilse Slootweg from Radboud University Medical Center for the annotation of fully delineated masks of prostate cancer for every bpMRI scan used in this study. This research is supported in parts by the European Union H2020: ProCAncer-I project (EU grant 952159) and Siemens Healthineers (CID: C00225450). Anindo Saha is also supported by an European Union: EACEA grant in the Erasmus+: Medical Imaging and Applications (MaIA) program.

%%Harvard
\bibliographystyle{model2-names.bst}\biboptions{authoryear}
\bibliography{refs}

\end{document}